# Developing a Theoretical Framework for Optofluidic Device Designing for System Identification in Systems Biology: the EGFR Study Case

Filippo MENOLASCINA[1] , Vitoantonio BEVILACQUA[1], Caterina CIMINELLI[1], Stefania TOMMASI[2] and Angelo PARADISO[2]
[1]*Polytechnic of Bari, Via E. Orabona 4, 70125, Italy*
[2]*Clinical Experimental Oncology Laboratory,National Cancer Institute, Via S. Hanhemann 10, 70126, Bari, Italy*

*f.menolascina@ieee.org*

**Abstract**: Identification of dynamics underlying biochemical pathways of interest in oncology is a primary goal in current systems biology. Understanding structures and interactions that govern the evolution of such systems is believed to be a cornerstone in this research. Systems theory and systems identification theory are primary resources for this task since they both provide a self consistent framework for modelling and manipulating models of dynamical systems that are best suited for the problem under investigation. We address herein the issue of obtaining an informative dataset $Z^N$ to be used as starting point for identification of EGFR pathway dynamics. In order to match experimental identifiability criteria we propose a theoretical framework for input stimulus design based on dynamical properties of the system under investigation. A feasible optofluidic design has been designed on the basis of the spectral properties of the driving inputs that maximize information content after the theoretical studies.

**Keywords:** systems biology, system identification, optofluidics, EGFR.

## 1. Introduction

Our understanding of molecular basis of complex diseases is being dramatically changed by systems investigation supported by the most advanced tools developed by the scientific community. In particular cancer has greatly benefited by systems level approaches since its development and progression is believed to be one of those system trajectories that arise from abnormal working states. The work by Hornberg and colleagues [5] pointed out the relevance of a systems biology approaches to the study of dynamics bringing to cancer. Epidermal Growth Factor Receptor pathway is one of those biochemical pathways that is believed to play a central role in cancer development. Even if comprehensive models of EGFR pathway still exist [6], many questions still remain open in terms of parameter identifiability and driving input design in systems biology. In the next paragraphs we give a brief but formal description the problem under investigation and we describe a theoretical framework thought to provide researchers with effective guidelines to optimal experimental design in systems biology.

## 2. Methods

Dynamic models of biological systems can be conveniently represented by a general nonlinear system $\Sigma_n$ in the state space form depending on a generally unknown parameter vector $p$:

$$\begin{cases} \dot{x}(t) = f(x(t),p) + \sum_{i=1}^{m} g_i(x(t),p)u_i(t) \\ y = h(u(t),x(t),p) \end{cases}$$

Where we have defined $x$ as an $n$-dimensional state variable in which we store variables that generally represent some form of energy in our system; $u$ is a $m$-dimensional vector that collects smooth input functions the system is supplied with; $y$ is the $r$-dimensional observed output and $p$ is a $v$-dimensional vector of unknown parameters. This notation also assumes that $f(\cdot)$, $g(\cdot)$ and $h(\cdot)$ are polynomial or rational functions of their parameters.

A priori *global identifiability* of the system-experiment model, is a condition guaranteeing theoretical uniqueness of solution to the problem of recovering the model parameters from input–output data. *Experimental identifiability*, on the other hand, is strictly tied to the feasibility of a general identification process that starts from a pre-assigned set of input-output data. This last criterion can be seen as a primary key to develop guidelines for optimal experimental design since obtaining *informative data*. Given a quasi-stationary data set $Z^\infty$ we say that $Z^\infty$ is informative enough with respect to the model $\mathcal{M}^*$ (the model to be identified) if, for any two models $W_1(q)$ and $W_2(q)$ in the set $\bar{E}[W_1(q) - W_2(q)z(t)]^2 = 0$ which implies that $W_1(e^{j\omega}) \equiv W_{12}(e^{j\omega})$ for almost all $\omega$. On the other hand a quasi-stationary data set $Z^\infty$ is informative if it is informative enough with respect to the model set $\mathcal{L}^*$consisting of all linear, time invariant (LTI) models. The concept of *informative dataset* is tightly related to concepts of *persistently exciting* inputs. This can be seen easily by observing that a quasi-stationary dataset $Z^\infty$ is informative if the spectrum matrix for $z(t) = [u(t) \;\; y(t)]^T$is strictly positive definite for almost all $\omega$. In fact, if we denote $W_1(q) - W_2(q) = \widetilde{W}(q)$ applying a well known theorem on signal filtration we can write: $\int_{-\infty}^{+\infty} \widetilde{W}(e^{j\omega})\Phi_z(\omega)\,\widetilde{W^T}(e^{-j\omega})\,d\omega = 0$, with $\Phi_x(\omega)$ spectrum of the signal $x(t)$ and where $\Phi_z(\omega) = \begin{bmatrix} \Phi_u(\omega) & \Phi_{uy}(\omega) \\ \Phi_{yu}(\omega) & \Phi_{yy}(\omega) \end{bmatrix}$. Since $\Phi_z(\omega)$ is positive definite, this implies that $\widetilde{W}(e^{j\omega}) \equiv 0$ almost everywhere that proves the previous statement. Moreover we can observe that, given the $\Phi_z(\omega)$, for the Shur's Lemma we can assure algorithm convergence only if $\Phi_u(\omega) > 0$ and $\Phi_u(\omega) - \Phi_{uy}(\omega)\Phi_{yy}(\omega)\Phi_{yu}(\omega) > 0$. Evidently the only block of this array we have control on is the one representing the spectrum of input signal which directly depends on dynamical properties of the driving input signal we design. It is therefore convenient to introduce the concept of *persistently exciting signal of order n* for a quasi-stationary stimulus $\{u(t)\}$: we say that a similar signal, with spectrum $\Phi_u(\omega)$ is persistently exciting of order n is, for all filters of the form $M_n(q) = m_1 q^{-1} + \cdots + m_n q^{-n}$ the relation $|M_n(e^{j\omega})|^2 \Phi_u(\omega) \equiv 0$ implies that $M_n(e^{j\omega}) \equiv 0$. Evidently the function $M_n(z)\, M_n(z^{-1})$ can have $n-1$ different zeros on the unit circle (since one zero is always at the origin) taking symmetry into account. Hence $u(t)$ is *persistently exciting of order n* if $\Phi_u(\omega)$ is different from zero on at least $n$ points in the interval $-\pi \le \omega \le \pi$. This is a direct consequence of the definition. Signals that show such properties have been investigated and include Pseudo Random Binary Signal, Generalized Binary Noise (or Random Binary Signal), Sum of Sines and Filtered Noise. Implementing such signals is quite simple from a computational point of view; however obtaining realizations of signals with such properties is an active area of research in current microfluidics. Developing geometries that satisfy physical conditions for the generation of signals compliant with the specifications imposed by the theoretical results we have shown is not a trivial task. Several alternative solutions have been proposed for signal modulation in microfluidic channels [1][2] [3]; they are based on diverse physical principles like boundary diffusion controlled by relative velocity (like in H filters), by exciting cells with diverse laminar flows that affect different parts of the cell etc. Here we propose a method based on the theory of Kelvin-Helmholtz discontinuities (the idea is presented in Fig. 1). The effect results from velocity shears between two fluid. Any time there is a non-zero curvature, the flow of one fluid around another will lead to a slight centrifugal force which in turn leads to a change in pressure thereby amplifying the ripple. The most familiar example of this is wind blowing over calm water. Tiny dimples in the smooth surface will quickly be amplified to small waves and finally to frothing white-caps. Any sort of surface tension will hinder KH instabilities. If there is some restoring force $T_g$, the instability will arise if $\Delta v^2 \geq \frac{2(\rho_1+\rho_1)}{\rho_1\rho_2}\sqrt{T_g(\rho_1-\rho_1)}$. On the other hand the stability requirement is met if $\Delta v^2 \leq 2v_A = \frac{2B}{\sqrt{\mu_o\rho}}$. Moreover we can control the frequency and wave number of the instabilities by analyzing the dispersion equation that states that $\frac{\omega}{k} = \frac{\rho U - \rho' U'}{\rho+\rho'} \pm \sqrt{\frac{g}{k}\frac{\rho-\rho'}{\rho+\rho'} - \frac{\rho\rho'(U-U')^2}{(\rho+\rho')^2}}$ with $\rho$ and $\rho'$ densities of the first and second fluid and $U$ and $U'$ velocities of the first and second fluid respectively.

## 3. Results and Discussion

According to all these specification we designed a microfluidic chip to be implemented using the polydimethilsiloxane (PDMS) technology. The mixing section of this chip is presented in Fig 2.

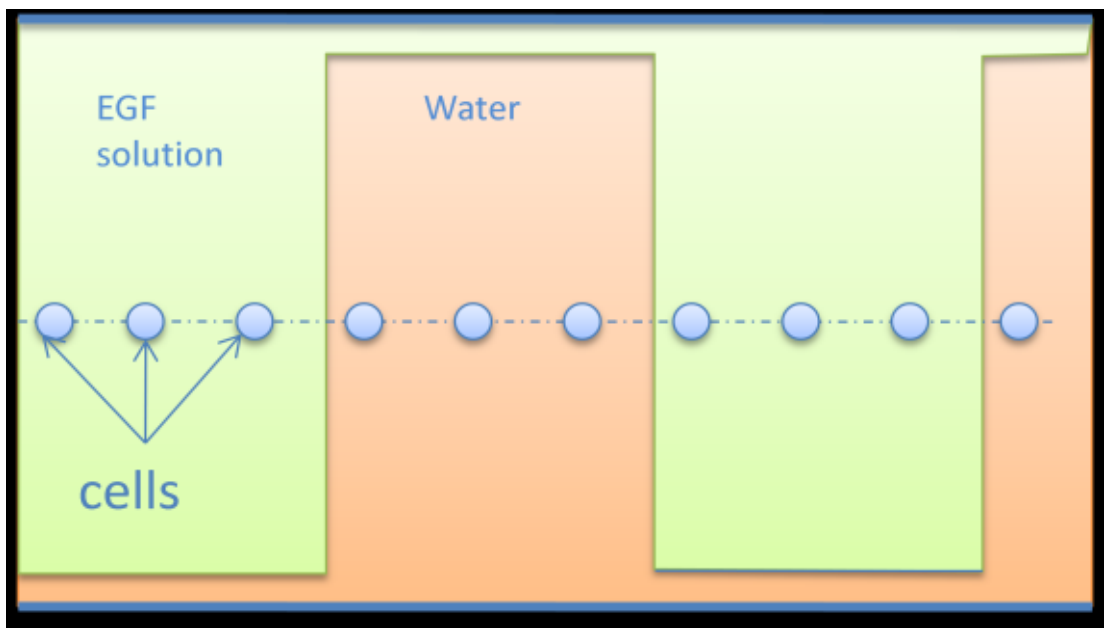

Fig. 1. Fluid behavior in the channel. Cells are stimulated by a modulated signal

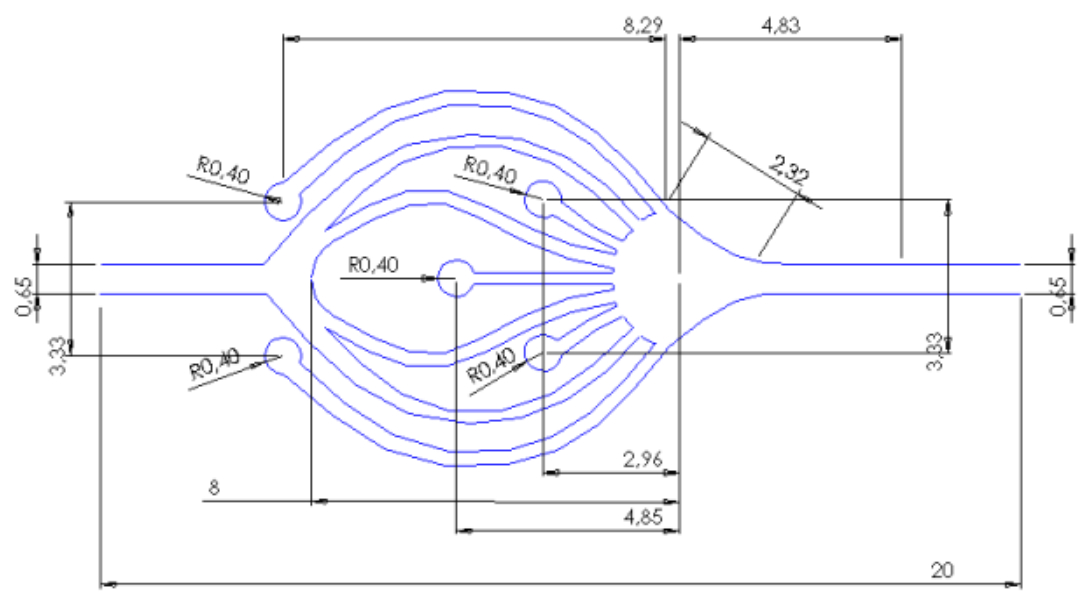

Fig. 2 Mixing section of the developed chip.

As it can be seen in the final part of this design it thought to let fluid converge in the common channel in which cells are anchored. Over this channel a here a fiber-coupled laser-diode module together with a photodetector allow to trace the presence and position of the cells; this ensures the . Computational fluid dynamics studies on this geometry carried out using ANSYS CFD showed acceptable behaviors even at low Reynold regimes. Tests carried out using Potterswheel [4] MATLAB package to test theoretical results for the EGFR pathway (reference model by Kitano et al [6]) confirmed the expectations about information richness of the obtained I/O dataset and pushed the interest for further research in this field.

## 4. References

[1] Takayama S, Ostuni E, LeDuc P, Naruse K, Ingber DE, GW. Whitesides. Selective Chemical Treatment of Cellular Microdomains Using Multiple Laminar Streams. Chem. Biol. 10(2): 123-130 (2003).

[2] Takayama S, Ostuni E, LeDuc P, Naruse K, Ingber DE, Whitesides GM. Subcellular positioning of small molecules. Nature 411:1016 (2001).

[3] Squires TM, Quake SR. "Microfluidics: Fluid physics at the nanoliter scale" *Rev. Mod. Phys.* July 2005, 7:3 pp. 977-1026.

[4] T. Maiwald, Potterswheel. http://www.potterswheel.de/

[5] J.J. Hornberg et al, Cancer: A Systems Biology disease, BioSystems 83 (2006) 81–90

[6] K Oda, Y Matsuoka, A Funahashi, H Kitano. A comprehensive pathway map of epidermal growth factor receptor signaling. - Mol Syst Biol, 25 May 2005